\documentclass[newstyle,twocolumn,proceedings]{rmaa}
\usepackage{rmaacite}

\title{Constraining the evolutionary histories of spiral disks}
\author{M. Moll\'{a} and E. Hardy 
 \affil{Universidad Aut\'{o}noma de Madrid,Cantoblanco (Spain)} \and
National Radio Astronomy Observatory 
(NRAO)\altaffilmark{1}}

\altaffiltext{1}{The National Radio Astronomy Observatory is a facility of
the National Science Foundation operated under cooperative agreement
by Associated Universities, Inc., U.S.A.}
\fulladdresses{
\item M. Moll\'{a}: Departamento de F\'{\i}sica Te\'{o}rica C-XI,
  Univerisdad Aut\'{o}noma de Madrid, Campus de Cantoblanco, 28049-Madrid
Spain (mercedes.molla@uam.es).
\item E. Hardy: National Radio Astronomy Observatory (NRAO), 
Casilla 36-D, Santiago, Chile (ehardy@nrao.edu).}

\shortauthor{ Moll\'{a} \& Hardy}
\shorttitle{Constraining the evolutionary histories}
\abstract{We study the old problem of the uniqueness of chemical
 evolution models. We showed in \scite{mol99} that multiphase models
 for three Virgo cluster galaxies were able to reproduce the observed
 radial distributions of spectral indices Mg2 and Fe52.  But those
 models may fit the present-epoch radial distributions with different
 star formation histories \cite{tos96}.  The two spectral indices
 above are in turn affected by the well known age-metallicity
 degeneracy which prevents the disentangling of age and metallicity
 for single stellar populations.  In this work we face both issues by
 analyzing a set of multiphase models for the Virgo galaxy NGC 4303.}

\resumen{Estudiamos el conocido problema de la unicidad de los modelos
de evoluci\'{o}n qu\'{\i}mica.  Hab\'{\i}amos demostrado en
\scite{mol99} que con los modelos multifase se pod\'{\i}an reproducir
las distribuciones radiales observadas de los \'{\i}ndices espectrales
Mg2 y Fe52 para tres galaxias de Virgo.  Dichos modelos, sin embargo,
pueden ajustar las distribuciones radiales observadas para el tiempo
presente con diferentes historias de formaci\'{o}n estelar
\cite{tos96}. Los dos \'{\i}ndices espectrales sufren a su vez la
llamada degeneraci\'{o}n edad-metalicidad, que impide discriminar al
mismo tiempo la edad y la metalicidad de las poblaciones estelares
simples. Trataremos ambas cuestiones analizando un conjunto de modelos
multifase para la galaxia de Virgo NGC~ 4303}

\listofauthors{M.~Moll\'{a} \& E.~Hardy}
\indexauthor{Moll\'{a}, M.}
\indexauthor{Hardy, E.}

\begin{document}
\maketitle

\section{Introduction}
\label{sec:intro}

In this work we compute a large number of multiphase chemical evolution
models for the Virgo galaxy NGC~ 4303. This model has been widely
described in \scite{fer94,mol99}. A protogalaxy is assumed to be a
spheroid composed of primordial gas with total mass $M(R)$. For NGC
4303 the rotation curves obtained by \cite{dis90} and \cite{guh88} are
used to compute $M (R)$. The galaxy is divided into concentric
cylindrical regions 1 kpc wide and the model calculates the time
evolution of the halo and disk components belonging to each one. The
halo gas falls into the galactic plane to form the disk, which is a
secondary structure in this scenario, at a rate $f$. This infall rate
$f$ is inversely proportional to the collapse time scale
$\tau_{coll}$, which, in turn, is assumed to depend on galactocentric
radius as: $\tau_{coll}= \tau_{0} \exp{(R-R0)/ \lambda_{D}}$, where
the characteristic time scale $\tau_{0}$ is determined using the total
mass of the galaxy, through the expression \cite{gal84}:
\begin{equation} 
\tau_{0}=\tau_{MWG}(\frac{M_{9,MWG}}{M_{9,gal}}){1/2}
\end{equation}

Stars form out in the halo, by a Schmidt law, while in the disk stars
form in two steps: molecular clouds $c$ form from the diffuse gas by a
Schmidt law, $c= \mu g^{1.5}$, then cloud-cloud collisions produce
stars by a spontaneous process, at a rate proportional to H: $SFR
\propto Hc{2}$. The efficiencies or probabilities to form molecular
clouds and stars, $\epsilon_{\mu}$ and $\epsilon_{H}$ respectively,
are assumed as dependent on morphological type T (see \pcite{mol02}).
The other two efficiencies corresponding the other processes described
are assumed constant for all halos and galaxies.

\section{The Computed Models}

We must now select these input parameters ($\tau_{0}$,$\lambda_{D}$
and T) for this galaxy. In order to take into account the
uncertainties in their selection, we choose them over a wide range of
values.  We thus run models with 5 values for the characteristics
collapse time scale: $\tau_{0}=$1, 4, 8, 12 and 16 Gyr. We also assume
5 possible values for the scale length: $\lambda_{D} = $1, 4, 8, 12
and 16 kpc. On the other hand, if we change the collapse time scale
values, the efficiencies --or equivalently T --must be changed
accordingly, in order to reproduce the observed radial distributions
which are the same used in \cite{mol99}.  We assume that T take values
from 1 to 20.  We thus run a total of 500 models with all possible
variations of these three parameters, $\tau_{0}$ , $\lambda_{D}$ and T.

\section{The $\chi^{2}$ optimization}

\begin{figure}
  \includegraphics[width=8cm]{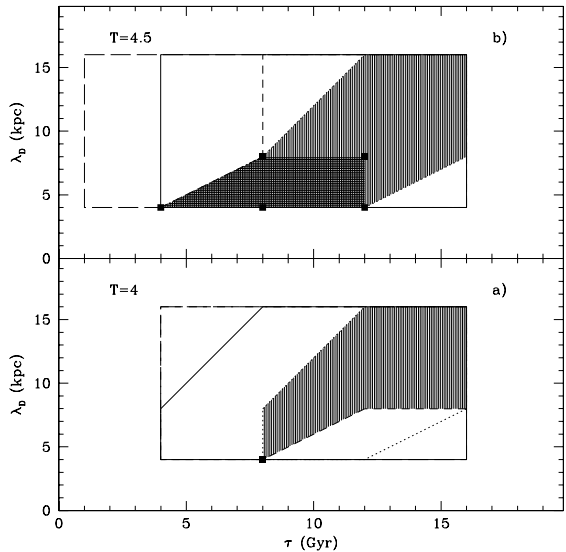}
  \caption{Probability contours representing regions with P$ \ge 97.5$ \%
 for all observational constraints for T$=4$ (panel a) and T$=4.5$
(panel b) as a function of $\tau$ and $\lambda$. }
\label{fig:param}
\end{figure}

We estimate the goodness of fit for these 500 models with the
statistical indicator $\chi^{2}$ (see \pcite{mol02a} for details).
First, by choosing those models able to fit
the present-time observational constraints with probabilities larger
than 97.5 \%, we find only 19 valid models among the 500 initial ones.

Next, radial distributions of Mg2 and Fe5270 are computed, by using
the chemical abundances and star formation histories predicted by the
above selected chemical evolutionary models as the necessary input for
the synthesis procedure which calculates spectral indices.  
\adjustfinalcols
We use again the statistical indicator $\chi^{2}$ to compare these
predicted spectral indices distributions with the data and thus to
select the adequate evolutionary histories over the disk.  Only 6
models are able to fit the spectral indices radial distributions with
the same level of confidence of 97.5\%.

We reduce, therefore, the possible star formation histories to those
predicted by the models whose parameters fall within a very restricted
range: $ 4 < T < 4.5$, $ 4 <
\tau_{0} < 12$ and $4 < \lambda_{D} < 8$. This may be clearly seen in
Fig.\ref{fig:param}, where the regions with probabilities higher
than 97.5\% for all observational constraints are shown as the zones
shaded with horizontal and vertical lines for $T=4.5$ (panel b), while
for $T=4$ (panle a) there is only one point 
($\tau_{0}=8$ Gyr, $\lambda_{D}=4$ kpc)

\section{Conclusions}
\begin{itemize}

\item When spectral indices are used these 19 possibilities reduce to
6. The spectrophotometrical information is essential to select the
adequate star formation histories.

\item Our technique of mixing both evolutionary chemical and synthesis
models proves a very useful tool which might be applied to other kind
of galaxies or galaxy regions.  

\item We suggest that radial measurements of
spectral indices on spiral disks must be consider seriously in the
preparation of observing programs.
\end{itemize}


\begin{thebibliography}

\bibitem[Distefano et al.{} <1990>]{dis90}
Distefano, A., Rampazzo, R., Chincarini, G., \& de Souza, R. 1990,
A\&AS, 86, 7

\bibitem[Ferrini et al.{} <1994>]{fer94}
Ferrini, F., Moll\'{a}, M., Pardi, C., \& D\'{\i}az, A. I. 1994, 
\apj, 427, 745 

\bibitem[Gallagher et al.{} <1984>]{gal84}
Gallagher, J. S., Hunter, D. A., \& Tutukov, A. V. 1984, \apj, 284, 544

\bibitem[Guhathakurta et al.{} <1988>]{guh88}
Guhathakurta, P., van Gorkom, J. H., Kotanyi, C. C., \& Balkowski,
C.1988, AJ, 96, 851

\bibitem[Moll\'{a} et al.{}<2002>]{mol02}
Moll\'{a}, M., D\'{\i}az, A. I., \& Ferrini, F. 2002, ApJ, submitted

\bibitem[Moll\'{a} \& Hardy <2002>]{mol02a}
Moll\'{a}, M., Hardy, E. 2002, AJ, accepted

\bibitem[Moll\'{a} et al.{} <1999>]{mol99}
Moll\'{a}, M., Hardy, E., \& Beauchamp, D.  1999, \apj, 513, 695 

\bibitem[Tosi <1996>]{tos96}
Tosi, M. 1996, in From Stars to Galaxies:
the impact of stellar physics on galaxy evolution, PASP Conference
Series, vol 98, Eds. C. Leitherer, U. Fritze-von Alvensleben, \&
J. Huchra, p.299 

\end{thebibliography}
\end{document}